\documentclass{PoS}

\usepackage{amssymb, amsthm, amsmath}
\usepackage{graphicx,graphics}
\usepackage{psfrag}
\usepackage{verbatim}
\usepackage{subfigure}
\usepackage{wrapfig}
\usepackage{amsfonts}

\title{Complex spectrum of spin models for finite-density QCD}

\ShortTitle{Complex spectrum of spin models for finite-density QCD}

\author{\speaker{Hiromichi Nishimura}\thanks{HN is supported by the RIKEN Special Postdoctoral Researchers Program.}\\
        RIKEN BNL Research Center\\
        E-mail: \email{hnishimura@bnl.gov}}

\author{Michael Ogilvie\\
        Washington University in St. Louis\\
        E-mail: \email{mco@physics.wustl.edu}}
        
\author{Kamal Pangeni\\
        Washington University in St. Louis\\
        E-mail: \email{kamalpangeni@wustl.edu}}

\abstract{We consider the spectrum of transfer matrix eigenvalues associated with Polyakov loops in lattice QCD at strong coupling. The transfer matrix at finite density is non-Hermitian, and its eigenvalues become complex as a manifestation of the sign problem. We show that the symmetry under charge conjugation $\mathcal{C}$ and complex conjugation $\mathcal{K}$ ensures that the eigenvalues are either real or part of a complex conjugate pair, and the complex pairs lead to damped oscillatory behavior in Polyakov loop correlation functions, which also appeared in our previous phenomenological models using complex saddle points. We argue that this effect should be observable in lattice simulations of QCD at finite density.}

\FullConference{34th annual International Symposium on Lattice Field Theory\\
		24-30 July 2016\\
		University of Southampton, UK}

\begin{document}

\section{Introduction}

Finite-density QCD is one of the class of theoretical models that has a sign problem: the partition function is a sum over complex weights
which cannot be interpreted as relative probabilities \cite{deForcrand:2010ys}. Many methods have been used in attempts to overcome the sign problem in finite-density QCD.
We have recently explored the implications of complex saddle points in phenomenological models of QCD at finite temperature and density \cite{Nishimura:2014rxa}. These models postulate effective potentials for the Polyakov loop $ \mathrm{tr}_{F}P_{x}$ and other
order parameters in such a way that the confinement-deconfinement transition of quenched QCD is incorporated. In all the cases studied in \cite{Nishimura:2014rxa}, a nonzero $\mu$ resulted in a complex saddle point for the eigenvalues of the Polyakov loop. A number of desirable results emerge from this. For example, the free energy is real at the complex saddle point and $\left\langle \mathrm{tr}_{F}P\right\rangle \ne\left\langle  \mathrm{tr}_{F}P^{\dagger}\right\rangle $. 
The mass matrix for Polyakov loops exhibits a new feature: the mass eigenvalues may form a complex conjugate pair, indicating the occurrence of spatially-modulated sinusoidal decay. Such behavior is forbidden by spectral positivity for $\mu=0$, but is possible when $\mu\ne 0$. 

We address the generality of this phenomenon by showing similar behavior in lattice QCD with static quarks in the strong-coupling
limit. We will use a transfer matrix formalism to determine the behavior of Polyakov loop correlation functions as a function of spatial separation \cite{Nishimura:2015lit}. These results are exact for any gauge coupling in $1+1$ dimensions, but also represent the leading-order result in the character expansion in higher dimensions. As the chemical potential of the static quarks is varied, we will show that there are large regions of parameter space where the eigenvalues of the transfer matrix form complex conjugate pairs, leading to damped oscillatory behavior of Polyakov loop correlation functions. The boundary of such a region in parameter space is referred to as a disorder line in condensed matter physics. The method used is completely different from the saddle point technique employed in \cite{Nishimura:2014rxa}, applied to a very different model, illustrating the generality of the behavior. Any reliable simulation method for finite-density lattice QCD should be able to reproduce our results, which thus can serve as a benchmark for the validation of algorithms.

\section{Setup}

$SU(N)$ pure gauge theory on $1+1$-dimensional lattice is exactly solvable.  Using the character expansion, we can integrate all the spatial links and write the partition function as
\begin{equation}
Z_{0}= \int \left[dP\right] \prod_{(x,y)} \sum_r c_r  \chi_r(P^{-1}_x) \chi_r (P_y) \equiv \int \left[dP\right] \prod_{(x,y)} T_0(x,y)
\end{equation}
where $[dP]$ is over Haar measure for the Polyakov loop $P_x$ on each spatial lattice site $x$, and $(x,y)$ is the nearest-neighbor pair. We choose the heat kernel action, for which the coefficients become $c_r = \exp(-m C_r) $, where $C_r$ is the quadratic Casimir invariant for the representation $r$, and $m$ is a mass parameter which depends on the gauge coupling and lattice spacing \cite{Menotti:1981ry}. Note that the transfer matrix associated with the Polyakov loop, $T_0(x,y)$, is in the spatial direction just as in spin systems in statistical mechanics.  

In order to keep the theory simple, we consider static quarks, which only move in a temporal direction \cite{Blum:1995cb}.  In this case, the transfer matrix becomes
\begin{equation}
T_s = T^{1/2}_0 \det \left[ 1+ z_1 P_x \right] \det \left[ 1+ z_2 P^{\dagger}_x \right] T^{1/2}_0
\label{T_s}
\end{equation}
where the determinants with the Polyakov lines $P_x$ and $P^\dagger_x$ are the quark and antiquark contributions at $x$ with the activities,
\begin{equation}
z_1 = e^{\left( \mu-M \right) /T}   \;\;\;\;\; \mbox{and} \;\;\;\;\; z_2 = e^{\left( -\mu-M \right) /T},
\end{equation}
respectively.
This particular form is chosen so that $T_s$ is Hermitian when $\mu=0$. While the transfer matrix corresponding to pure gauge fields, $T_{0}$,
is Hermitian, the final transfer matrix, $T_s$, that includes the effect of static quarks and antiquarks is no longer Hermitian. $\mathrm{tr}_F P$ and $\mathrm{tr}_F P^{\dagger}$ connect between different representations so they introduce off diagonal elements in the transfer matrix. Since $\mathrm{tr}_F P$ and $\mathrm{tr}_F P^{\dagger}$are different for $\mu\ne0$, the off-diagonal elements are no longer symmetric and the transfer matrix is non-Hermitian. Therefore the eigenvalues of the transfer matrix are in general complex. 

For very heavy quarks, the transfer matrix becomes \cite{Ogilvie:2008zt}
\begin{equation}
T_h = T^{1/2}_0 \exp \left[z_1\mathrm{tr}_F P \right] \left[z_2 \mathrm{tr}_F P^\dagger \right] T^{1/2}_0,
\label{T_h}
\end{equation}
which is an approximation of $T_s$ in the limit $z_1, z_2 \ll 1$ and is the lowest order in the activities of static fermions and bosons. It therefore misses the effects of Pauli blocking as well as the symmetries of the fermion determinant, which will be discussed in the next section. Nevertheless, we find that the eigenvalues of the transfer matrix also becomes complex in some region of the parameter space.  
At very strong coupling, the Hamiltonian takes the form
\begin{equation}
H = - J \sum_{(i,j)} \mbox{Re} \left[\mathrm{tr}_F P_i \mathrm{tr}_F P^\dagger_j \right] - h_R \sum_i \mbox{Re} \left[\mathrm{tr}_F P_i \right]- i h_I \sum_i \mbox{Im} \left[\mathrm{tr}_F P_i \right] .
\end{equation}
It is the Potts model with the complex magnetic field for the case of $Z(3)$, which was first studied in \cite{DeGrand:1983fk}.  

In $1+1$ dimensions, the results obtained from the transfer matrix are exact at any value of the coupling. In the strong coupling region, the results from $1+1$ dimensions are also valid in higher dimensions to leading order in the character expansion. This result was noted long ago in \cite{Kogut:1981ny}.
In any dimension, the strong-coupling expansion is a convergent expansion with a finite radius of convergence, so there will be some region around $g^{-2}=0$ where the lowest order result is a good approximation. 

\section{$\mathcal{CK}$ symmetry}

Although the eigenvalues of the non-Hermitian transfer matrices, (\ref{T_s}) and (\ref{T_h}), can be in general complex, the global symmetry of finite-density QCD restricts the form of eigenvalues as we now discuss. Our models inherit from finite-density QCD invariance under the combined action of charge conjugation $\mathcal{C}$ and complex conjugation $\mathcal{K}$  \cite{Nishimura:2014rxa}. 
Complex conjugation is an antilinear symmetry, changing not only fields but also complex-conjugating ordinary numbers. 
It is easy to see that the combined effect of $\mathcal{CK}$ is to leave the action invariant, so the transfer matrix $T$ commutes with $\mathcal{CK}$,
\begin{equation}
\left[T, \mathcal{CK} \right] =0 .
\end{equation}
From this, it can be shown that all the eigenvalues of the transfer matrix are either real or are part of a complex conjugate pair. 
As a consequence, correlation functions of operators that couple to eigenstates of $T$ with complex eigenvalues will generally exhibit some amount of sinusoidally-modulated exponential decay rather than the usual exponential decay found in models with Hermitian actions \cite{Nishimura:2014rxa,Ogilvie:2008zt}. We will show below that this sinusoidal modulation is present in strong-coupling
QCD with a finite density of both static and heavy quarks. 

In our model, $\mathcal{CK}$ symmetry has a physical interpretation.  As always, the charge conjugation $\mathcal{C}$ is a particle-antiparticle transformation, and it is explicitly broken when the chemical potential is nonzero.  The complex conjugation $\mathcal{K}$ can be here interpreted as the particle-hole transformation because $\mathcal{K}$ operation,
\begin{equation}
\det\left[1+z_{1}P_{x}\right] \rightarrow \det\left[1+z_{1}P^*_{x}\right]  = z_{1}^{N}\det\left[1+z_{1}^{-1}P_{x}\right],
\end{equation}
is equivalent to the particle-hole transformation, $z_1 \rightarrow 1/z_1$, up to the factor $z^N_1$.
This factor of $z_{1}^{N}$ represents the Boltzmann factor for a completely filled state at a site. Although the factor does contribute to the
free energy, it does not affect expectation values. 
This operation therefore reflects the equivalence between particles and holes:
a particle (relative to the vacuum) is equivalent to $N-1$ holes (relative
to the completely filled state) at the same site. In the special case
where $z_{2}=0$, there is an exact particle-hole symmetry under $z_{1}\rightarrow1/z_{1}$. At this point, the transfer matrix is hermitian; this is precisely the absence of a sign problem at ``half-filling" for static quarks. See \cite{Rindlisbacher:2015pea} for an extensive treatment of this property.  
This extends to an approximate particle-hole symmetry when antiparticle
effects are small. If we apply the above identity to antiparticles as well as particles,  we obtain an exact symmetry of the complete theory:
The model is invariant under the transformation $\left(z_{1},z_{2}\right)\rightarrow\left(1/z_{2},1/z_{1}\right)$.
Note however that  this transformation takes the physical region $z_1, z_2 < 1$ into the unphysical region $z_1, z_2 >1$.

\section{Results}

\begin{figure}
\hfill
\begin{minipage}[t]{.43\textwidth}
\begin{center}
\includegraphics[width=0.8\textwidth]{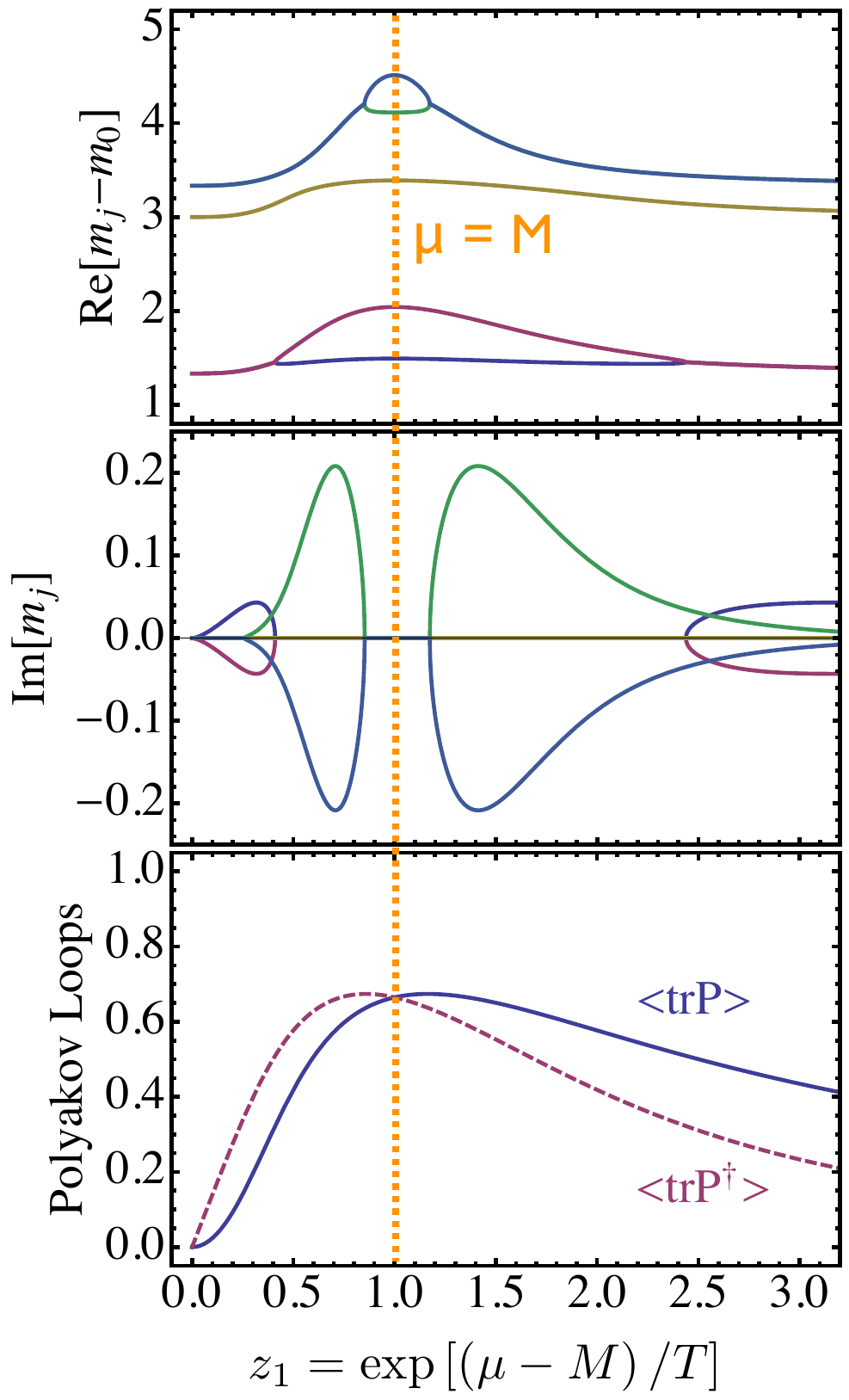}
\caption{Mass spectrum and Polyakov loops for the case of static quarks in the heavy-dense limit.}
\label{Mass_Spectrum}
\end{center}
\end{minipage}
\hfill
\begin{minipage}[t]{.55\textwidth}
\begin{center}
\includegraphics[width=0.8\textwidth]{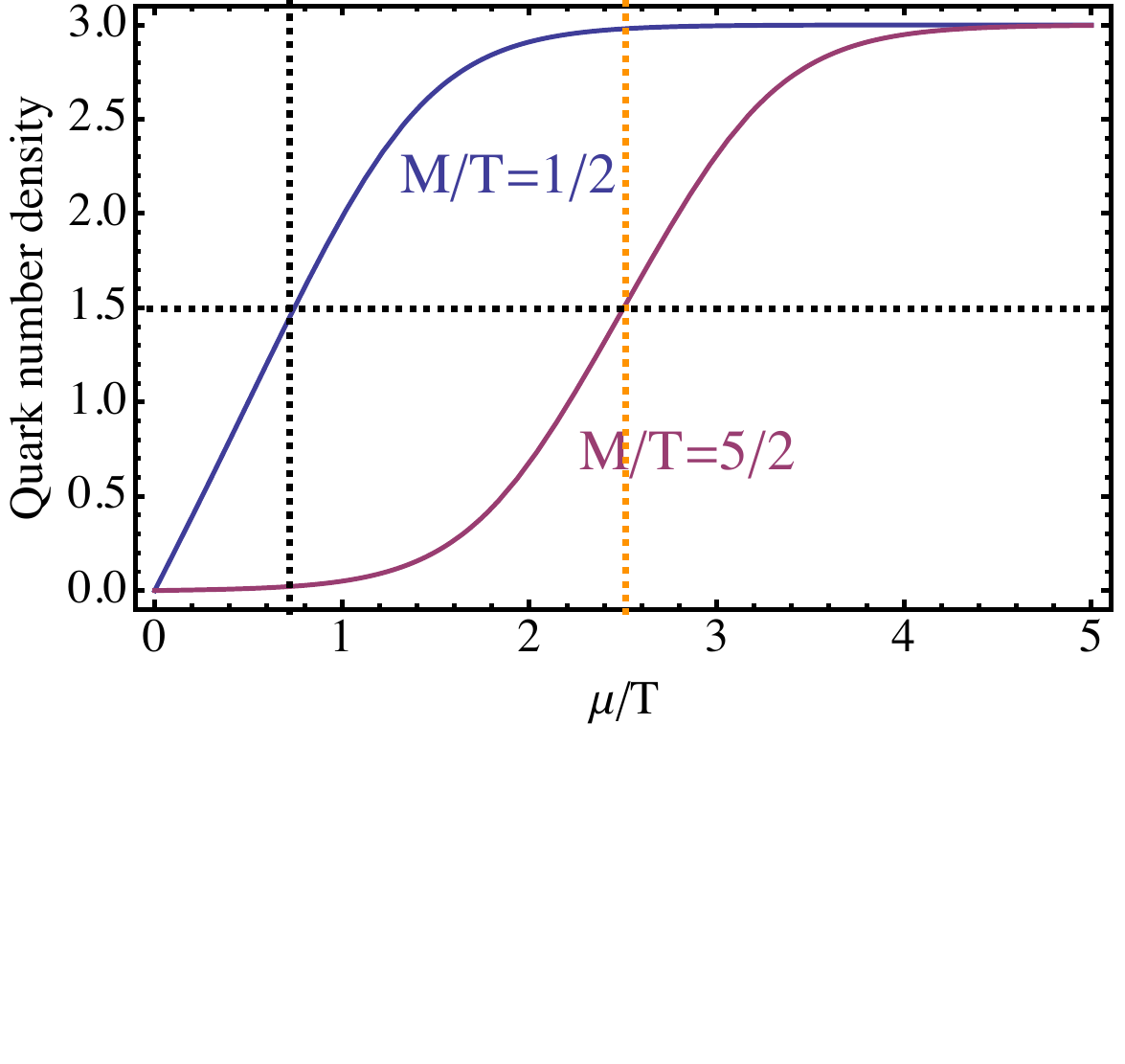}
\caption{Quark number density for the case of static quarks.}
\label{Number_Density}
\end{center}
\end{minipage}
\hfill
\end{figure}

We first consider the case of static quarks in the heavy-dense limit where the effects of antiquarks are neglected. We diagonalize the transfer matrix to write $T_s = \mbox{diag} (e^{-m_0 a},e^{-m_1 a}, \dots)$.
Figure \ref{Mass_Spectrum} shows the real and imaginary parts of the mass spectrum in the unit of lattice spacing, $\mbox{Re} \left[m_j-m_0 \right]$ and $\mbox{Im} \left[m_j\right]$, for low-lying eigenstates when the parameter $m$ in $T_0$ is set to be one. We expect that the ground state energy $m_0$ is real for any $\mu$ because there should be no spontaneous symmetry breaking of $\mathcal{CK}$.  We have numerically checked this. The figure also shows Polyakov loop expectation values. As seen in the plots, the masses start out real for $z_{1}=0$ but quickly take on complex values for non-zero $z_{1}$. As $z_{1}$ increases, the magnitude of complex part of the mass gradually increases before dropping back to zero.  The plots clearly reflect the particle-hole symmetry under $z_{1}\rightarrow1/z_{1}$. 
The real part of mass spectrum is highest when $\mu=M$.
This point is special because the theory is Hermitian at half-filling and the mass spectrum must be real.
Furthermore, there appears to be a region around $z_{1}=1$ where
the mass spectrum is real. The size of this region is largest for
the lowest representations. 
For $\mu<M$, $\left\langle \mathrm{tr}_{F}P\right\rangle $ is less than $\left\langle 
\mathrm{tr}_{F}P^{\dagger}\right\rangle $, implying that the free energy cost of inserting a fermion into the system is greater than that of inserting an antifermion. 
For $\mu>M$, this behavior is reversed in accordance with the $z_{1}\rightarrow1/z_{1}$ symmetry. At $\mu=M$, the two expectation values are equal. 

The quark number density for the case of static quarks is shown in Fig.~\ref{Number_Density} as a function of $\mu/T$ for $M/T=1/2$ and $5/2$; the parameter
$m$ is set to $1$. The most obvious feature is the saturation of the number density at $3$ for large $\mu$. In the context of lattice gauge theories at finite density, saturation was first discussed in \cite{Hands:2006ve} for the case of $SU(2)$, where there is no
sign problem. See \cite{Rindlisbacher:2015pea,Langelage:2014vpa} for recent discussions of saturation effects in strong-coupling models of $SU(3)$ at finite density. Saturation is also observed in recent Langevin simulation with heavy quarks \cite{Aarts:2016qrv}. For the heavier quark mass $M/T=5/2$, we expect that antiquark effects are negligible for $\mu\simeq M$, and the system has an approximate particle-hole
symmetry at $M=\mu$. This in turn implies that the quark number density reaches half-filling $(1.5)$ at $M=\mu$. As may be seen from the figure, that expectation is confirmed. For the lighter quark mass, $M/T=1/2$, antiquark effects are not negligible when $\mu\simeq M$, and antiquark contributions lower the number density at $\mu=M$ below the half-filling value. In both cases, the number density saturates at some value of $\mu$ larger than $M$. 

\begin{figure}
\begin{center}
\subfigure[$\left<\mathrm{tr}_F P^\dagger(r) \mathrm{tr}_F P(0) \right>$]{\includegraphics[width=2.9in]{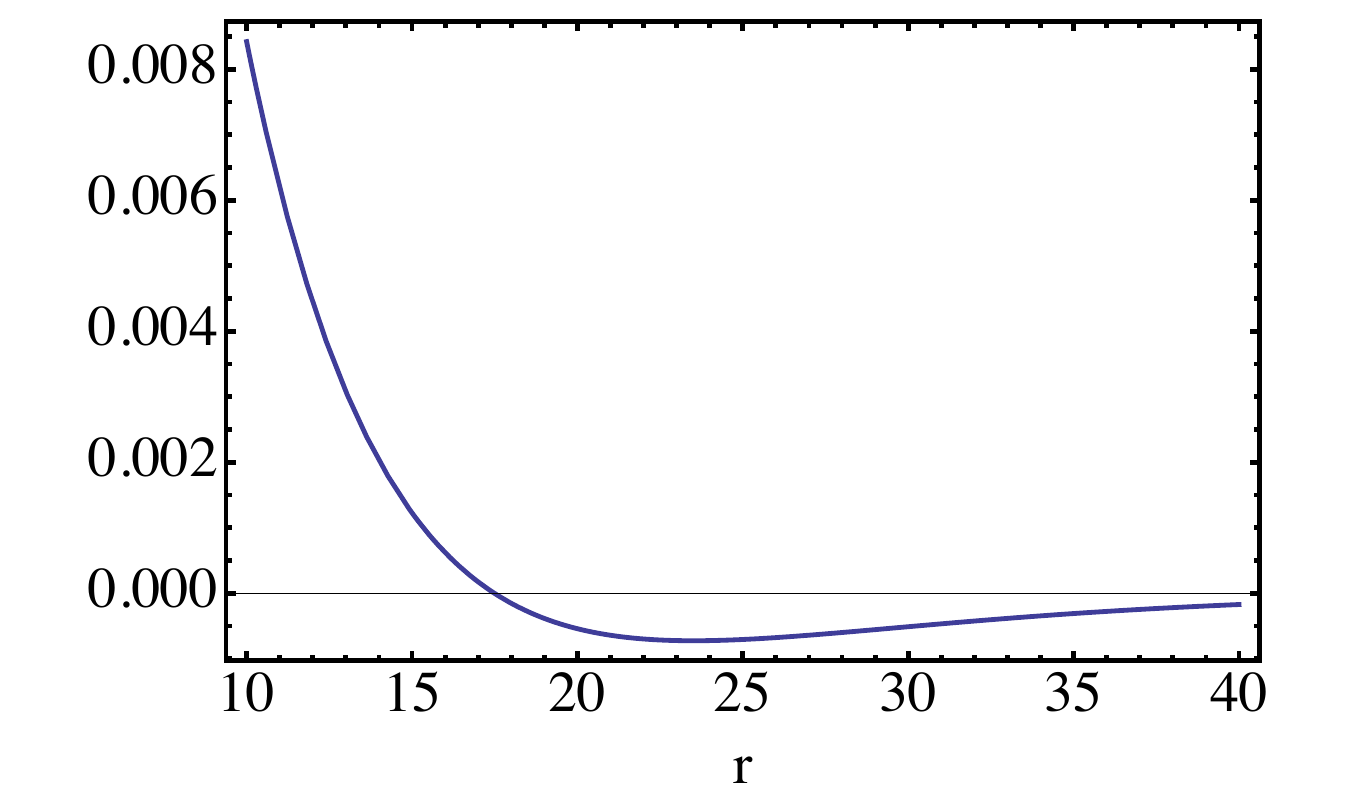}\label{Polyakov_Correlator}}
\subfigure[Oscillatory part of $\left<\mathrm{tr}_F P^\dagger(r) \mathrm{tr}_F P(0) \right>$]{\includegraphics[width=2.9in]{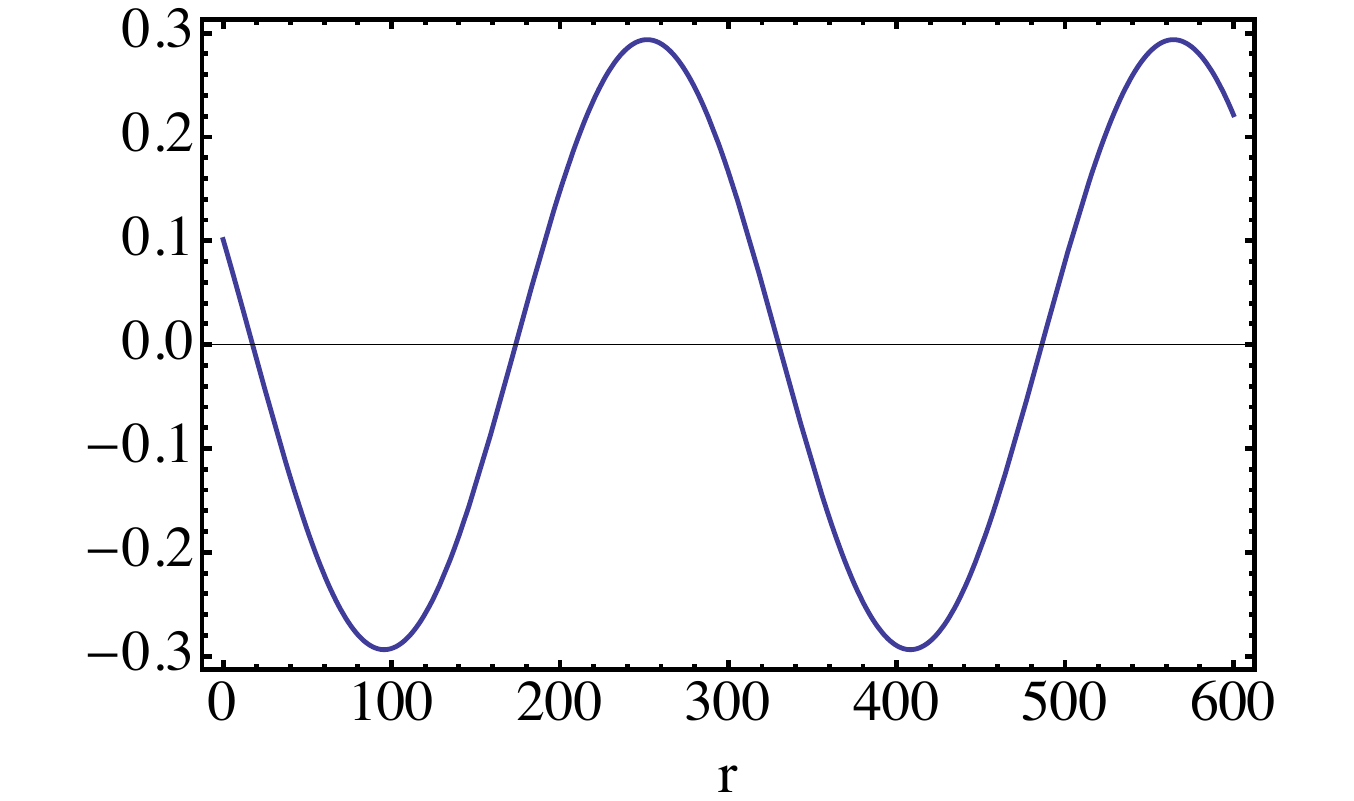}\label{Polyakov_Oscillatory}}
\vspace{-2mm}
\caption{Sinusoidal oscillation of the Polyakov loop correlator for the case of static quarks.} \label{fig1}
\end{center}
\end{figure}

As may be seen from Fig.~\ref{Mass_Spectrum}, the magnitude of the imaginary part of any eigenvalue is generally substantially smaller than the real part and may be difficult to observe directly in simulations.  Nevertheless, it may be possible to observe the modulated decay directly in some circumstances. Figure \ref{Polyakov_Correlator} shows the $3-\bar{3}$ Polyakov loop correlation
function $\left\langle \mathrm{tr}_{F}P^{\dagger}\left(r\right) \mathrm{tr}_{F}P\left(0\right)\right\rangle $
as a function of distance $r$ for $m=0.1$ and $z_{1}=0.08$ with $z_{2}=0$.
The clear minimum at $r\simeq22$ is a consequence of sinusoidal modulation
and reminiscent of the behavior of density-density correlation functions
in liquids. Note also that the correlation function drops below zero,
also as a consequence of the spatial modulation. Figure \ref{Polyakov_Oscillatory} shows the oscillatory part of Polyakov loop by taking out the exponential decay by hand.   

\begin{figure}
\begin{center}
\subfigure[Static quarks]{\includegraphics[width=2.4in]{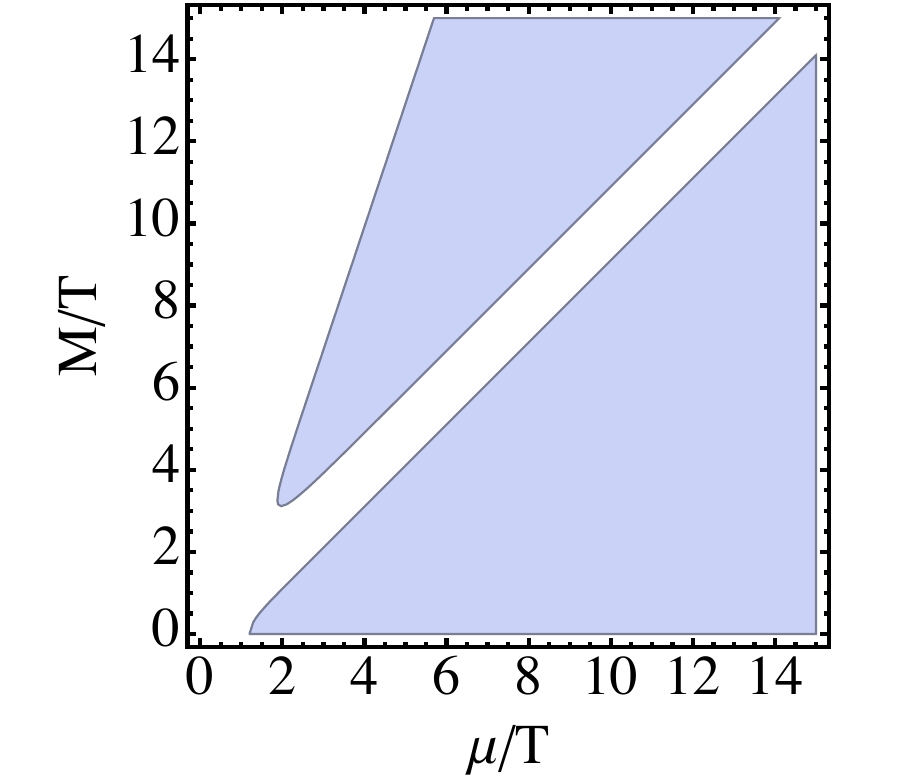}\label{DL_Static}}
\subfigure[Heavy quarks]{\includegraphics[width=2.4in]{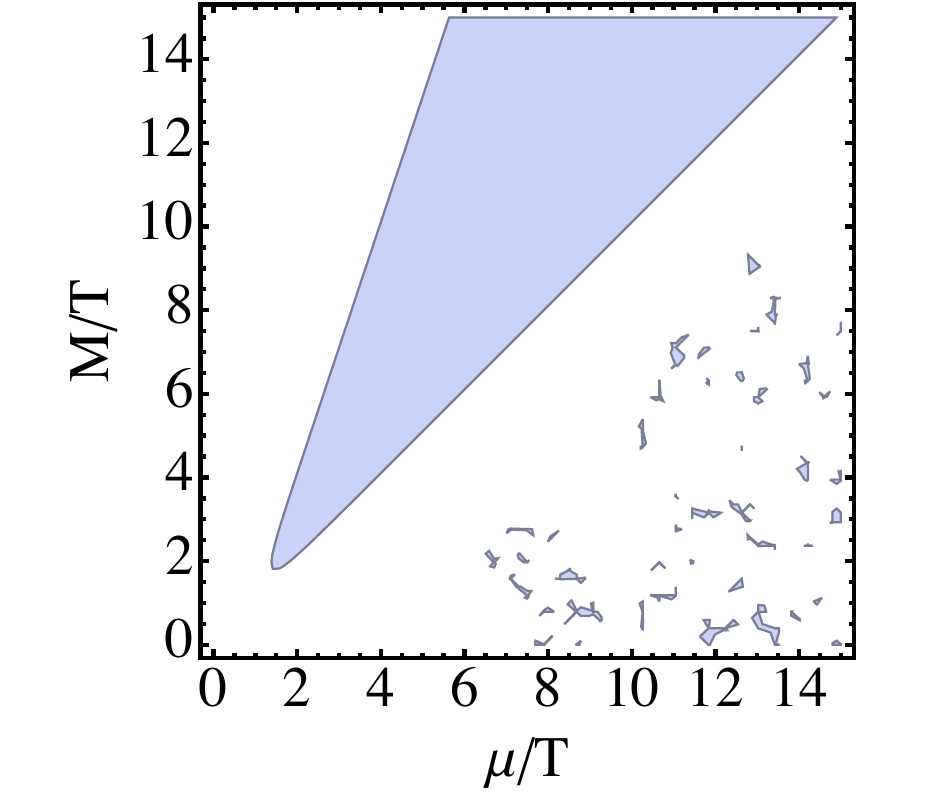}\label{DL_Heavy}}
\vspace{-2mm}
\caption{The disorder lines for the static quarks and the heavy quarks.} \label{DL}
\end{center}
\end{figure}

Figure \ref{DL} shows the region in the $\mu-M$ plane where the low-lying eigenvalues form a complex conjugate pair, i.e., the disorder line. The parameter $m$ is set to $1$ for both cases.  For the case of static quarks as shown in Fig.~\ref{DL_Static}, we see a symmetric structure near $\mu = M$ when the mass is sufficiently large and so the heavy-dense limit is valid. As a result, there are two regions of complex mass separated by the hermitian point $\mu=M$.  For the case of heavy quarks as shown in \ref{DL_Heavy}, the complex region for $\mu < M$ is similar, but it goes away as we go beyond the hermitian point of $\mu=M$, because there is no particle-hole symmetry in this case.  
 
\section{Conclusions}

The spatial transfer matrix associated with Polyakov loops in finite-density QCD with static quarks (\ref{T_s}) and heavy quarks (\ref{T_h}) have complex eigenvalues over a significant
region of parameter space in strong-coupling limit. The appearance of complex eigenvalues is a direct consequence of the non-hermiticity
of the transfer matrix. This is a manifestation of the sign problem in finite-density QCD. 
The invariance of finite-density QCD under $\mathcal{CK}$ symmetry ensures that the eigenvalues are either real or part of complex conjugate
pair. The complex conjugate pairs in turn give rise to sinusoidal
modulation of Polyakov loop correlation function. We have seen in Fig.~(\ref{DL}) that both models, (\ref{T_s}) and (\ref{T_h}), show this behavior in some regions of the parameter space. 

The occurrence of conjugate complex mass pairs and sinusoidal modulation
of Polyakov loop correlation functions was previously observed by
us in the study of phenomenological models of QCD using a saddle point
approximations \cite{Nishimura:2014rxa}. The presence
of similar phenomenon in lattice models of QCD at strong coupling
strongly suggests the generality of the phenomenon. It has been suggested
\cite{Patel:2011dp}
that sinusoidal modulation of this type might be observed in lattice
simulations and heavy ion experiments. In both phenomenological models
and in lattice strong-coupling calculations, the imaginary part of
the mass is significantly smaller than the real part, suggesting that
the direct observation of modulation might be difficult because of
a long wavelength.
Recently the case of a $Z(3)$ spin model with complex external fields has also been studied using an extended form of mean field theory \cite{Akerlund:2016myr}, with conclusions similar to ours.
 However, there is another way to observe the splitting
of the spectrum into complex pairs in lattice simulations. Suppose
that the imaginary part of the masses are too small to be directly
observed and can be neglected. In the regions where there are complex
conjugate eigenvalue pairs, the real parts of the eigenvalues are
degenerate, but outside of those regions they are different. This
effect should be present and observable in lattice simulations of
finite-density QCD, and may provide a strong test for finite-density
algorithms.

\end{document}